\documentclass[11pt]{article}

\usepackage{amsfonts}
\usepackage{amsbsy}
\usepackage{epsfig}
\usepackage{latexsym}
\input amssym.def
\input amssym.tex

\def\bequ{\begin{equation}}
\def\eequ{\end{equation}}
\def\barr{\begin{array}}
\def\earr{\end{array}}

\def\ben{\begin{equation}}
\def\een{\end{equation}}
\def\bena{\begin{eqnarray}}
\def\eena{\end{eqnarray}}

\newcommand{\sect}[1]{\setcounter{equation}{0}\section{#1}}

\advance \topmargin by -\headheight
\advance \topmargin by -\headsep
\evensidemargin \oddsidemargin
\advance\hoffset by -3mm  
\advance\voffset by  8mm  
\def\spa#1{\phantom{\fbox{\rule[-#1cm]{0cm}{0cm}}}}

\begin{document}
\hfuzz=100pt
\title{{\Large \bf{The Kerr-Newman-G\"odel Black Hole}}}
\author{\\C A R Herdeiro\footnote{E-mail:crherdei@fc.up.pt}
\\
\\
Centro de F\'\i sica do Porto,
\\ Faculdade de Ci\^encias da Universidade do Porto,
\\ Rua do Campo Alegre, 687, 4169-007 Porto, Portugal.}

\date{July 2003}
\maketitle


\begin{abstract}
By applying a set of Hassan-Sen transformations and string
dualities to the Kerr-G\"odel solution of minimal $D=5$
supergravity we derive a four parameter family of five dimensional
solutions in type II string theory. They describe rotating,
charged black holes in a rotating background. For zero background
rotation, the solution is $D=5$ Kerr-Newman; for zero charge it is
Kerr-G\"odel. In a particular extremal limit the solution
describes an asymptotically G\"odel BMPV black hole.

\end{abstract}

\sect{Introduction} Two particularly interesting solutions of
minimal supergravity in $D=5$ are the maximally supersymmetric
G\"odel type universe \cite{Gauntlett:2002nw} and the half
supersymmetric BMPV black hole \cite{Breckenridge:96}. The former
is a homogeneous space with Closed Timelike Curves through every
point in spacetime. Such Closed Timelike curves can be made to
vanish by uplifting the solution to higher dimensions
\cite{Herdeiro:02}, the reason being that these G\"odel type
universes are U-dual to pp-waves \cite{Boyda:2002ba}. This fact makes
the G\"odel type universe even more interesting, due to the
possibility of quantizing strings in this background and its
relation to a certain limit of Super Yang Mills
\cite{Berenstein:2002jq}. On the gravitational side the pp-waves dual
to the G\"odel universe arise as Penrose limits of (D-brane configurations) near horizon geometries.

The BMPV black hole is the only asymptotically flat, rotating,
supersymmetric black hole with a regular, finite size horizon and
thus finite entropy, known in string theory. See \cite{Reall:2002bh} for a uniqueness theorem concerning this solution. Its existence is made
possible due to the particular Chern-Simons coupling of $D=5$
minimal supergravity \cite{Gauntlett:1998fz} and the fact that the spacetime
dimension allows imposing a self-duality condition on the exterior
derivative of the rotation one-form \cite{Herdeiro:00}. Embedded
in type IIB string theory it corresponds to a deformation of the
$D1$-$D5$ system for which one can compute the degeneracy of
states at high level and find agreement with the classical
Bekenstein-Hawking formula. On the string theory side angular
momentum corresponds to R-charge carried by the quantum states \cite{Breckenridge:96}.
The BMPV spacetime can only be interpreted as a black hole for
sufficiently small values of angular momentum; otherwise there is
no horizon and the spacetime becomes a non-singular repulson
\cite{Gibbons:99} with Closed Timelike Curves passing through any
point. On the string theory side unitarity is violated when this happens \cite{Herdeiro:00}.

In this letter we will show that these two remarkable spacetimes
can be superimposed as a solution to type II string theory; there
is a BMPV-G\"odel black hole. That black holes can exist in the
G\"odel universe, was shown in \cite{Herdeiro:02}, where an
extreme Reissner-Nordstrom black hole was discussed. Subsequently
Gimon and Hashimoto \cite{Gimon:2003ms} constructed a Kerr black
hole which is asymptotically G\"odel. We will use their result as
the starting point for a well known procedure of generating
charged black holes in string theory. Thus we will generate a
Kerr-Newman-G\"odel black hole. In the extremal limit this yields
the advertised BMPV-G\"odel solution. 

The G\"odel `deformation' of the BMPV black hole becomes irrelevant close to the horizon.
Thus, the near horizon geometry and the entropy are unchanged from the asymptotically flat case.
In AdS/CFT this means that the background rotation is a sub-leading effect in $\alpha'$, in the decoupling limit \cite{Herdeiro:02}.

This spacetime is not a solution of $D=5$ minimal supergravity, which explains why it was not found in \cite{Gauntlett:2002nw}, but it reduces to 
a solution of this theory both for vanishing `mass' and for vanishing background rotation, when it yields the G\"odel and BMPV backgrounds, respectively. 
In general, our solution can be characterized as a type IIA solution on $K3\times S^1$, although it can certainly be dualized to yield a heterotic
or type IIB solution.

From the viewpoint of the BPS spectrum of string theory this is qualitatively a new state, which from the five dimensional gravitational
viewpoint includes both asymptotically decaying and asymptotically dominating `angular momentum' or, more accurately, inertial frame dragging effects.

\sect{The $D=5$ Kerr-Newman and Kerr-G\"odel}
The minimal supergravity theory in five spacetime dimensions was constructed in \cite{cremmer,Chamseddine:1980sp}. We take the action to be
\bequ
{\mathcal{S}}=\frac{1}{16\pi G_5}\int d^5 x\sqrt{-g}\left(R-F^2-\frac{2}{3\sqrt{3}}\tilde{\epsilon}^{\alpha \beta \gamma \mu \nu}F_{\alpha \beta}F_{\gamma \mu} A_{\nu}\right) \ , \label{5dsugra}\eequ
where $F=dA$, $\tilde{\epsilon}$ is the Levi-Civita tensor, related to the Levi-Civita tensor density by $\tilde{\epsilon}^{\alpha \beta \gamma \delta \mu}=\epsilon^{\alpha \beta \gamma \delta \mu}/\sqrt{-g}$ and we use a `mostly plus' signature. This theory admits as an uncharged solution the five dimensional Kerr; it is
parameterized by three parameters (mass and two angular momenta):
$m, a, b$. The metric reads \cite{Myers:86}

\bequ \barr{l} \displaystyle{
ds^2=-dt^2+\frac{r^2\Delta dr^2}{(r^2+a^2)(r^2+b^2)-2mr^2}+\Delta
d\theta^2+\sin^2{\theta}(r^2+a^2)d\phi^2+\cos^2{\theta}(r^2+b^2)d\psi^2}\spa{0.5}\\
\displaystyle{~~~~~~~~
+\frac{2m}{\Delta}\left[dt-(a\sin^2{\theta}d\phi+b\cos^2{\theta}d\psi)\right]^2}
\ , \label{Kerrab} \earr \eequ
and the gauge potential is zero.  The function $\Delta$ is defined as
\bequ
\Delta\equiv r^2+a^2\cos^2\theta+b^2\sin^2\theta \ . \eequ For $m=0$ we
have Minkowski space. If $a=b=0$ the spatial part is
${\mathbb{R}^4}$ written in spherical bipolar coordinates. That is, we take
two orthogonal 2-planes in ${\mathbb{R}^4}$ each parameterized by
an azimuthal coordinate ($\phi$ and $\psi$ each with range
$2\pi$); $\theta$ is the polar coordinate between the 2-planes
with range $\pi/2$. If $a,b\neq 0$, the spatial part is written in
oblate bipolar coordinates, which are adequate for describing
squashed 3-spheres rather than round ones. In this Kerr metric,
all squashing is contained in the one-form \bequ
a\sin^2{\theta}d\phi+b\cos^2{\theta}d\psi=\frac{a+b}{4}\sigma^3_L+\frac{b-a}{4}\sigma^3_R
\ .  \label{rot1f} \eequ We have introduced a particular left
($\sigma_L^3$) and right ($\sigma_R^3$) one-form of SU(2) for which we
follow the conventions of \cite{Herdeiro:02}.\footnote{The
relation between the bipolar coordinates above
($\theta,\phi,\psi$) and the Euler angles
($\tilde{\theta},\tilde{\phi},\tilde{\psi}$) used in
\cite{Herdeiro:02} is $\tilde{\theta}=2\theta$,
$\tilde{\phi}=\phi+\psi$, $\tilde{\psi}=\psi-\phi$.} Explicitly, in terms of the bipolar coordinates used in (\ref{Kerrab}), \bequ
\sigma_L^3=2(\cos^2{\theta}d\psi+\sin^2{\theta}d\phi) \ , \ \ \ \ \ \sigma_R^3=2(\cos^2{\theta}d\psi-\sin^2{\theta}d\phi)\ . \eequ

One thinks of $a (b)$ as the parameter associated with angular
momentum in the $\phi (\psi)$ 2-plane, $J_{\phi} (J_{\psi})$. One
the other hand, since spherical symmetry in $4+1$ dimensions is an
$SO(4)\simeq SU(2)_L\times SU(2)_R$ symmetry, it is natural to
define left, $J_L$, and right, $J_R$, angular momentum as the
Casimir operators of $SU(2)_L$ and $SU(2)_R$. The presence of
$\sigma^3_L$ $(\sigma^3_R)$ in (\ref{rot1f}) turns on $J_L$
($J_R$). Thus we can roughly say that \bequ J_L=J_{\psi}+J_{\phi}
\ , \ \ \ \ \ \ J_R=J_{\psi}-J_{\phi} \ . \eequ

The Kerr solution with either $J_L=0$ or $J_R=0$ simplifies dramatically. Defining a new radial coordinate
$R^2\equiv \Delta=r^2+a^2$, the `Left Kerr'  ($a=b$) and `Right Kerr' ($a=-b$) are written as \bequ
ds^2=-dt^2+\frac{R^4dR^2}{R^4-2m(R^2-a^2)}+\frac{R^2}{4}\left((\sigma^1_{L,R})^2+(\sigma^2_{L,R})^2+
(\sigma^3_{L,R})^2\right)+\frac{2m}{R^2}\left[dt\pm\frac{a}{2}\sigma_{L,R}^3\right]^2
\ . \label{leftkerr}\eequ
We use the convention that upper (lower) signs refer to the right (left) solution. Of course Right Kerr and Left Kerr are the same solution; explicitly they are interchanged by
\bequ
a\rightarrow -a \ , \ \ \ \ \phi\rightarrow -\phi \ . \eequ
Note that in bipolar coordinates
\bequ
(\sigma^1_L)^2+(\sigma^2_L)^2+(\sigma^3_L)^2=(\sigma^1_R)^2+(\sigma^2_R)^2+
(\sigma^3_R)^2=4(d\theta^2+\sin^2{\theta}d\phi^2+\cos^2{\theta}d\psi^2) \ . \eequ

The five-dimensional Left and Right Kerr-Newman have an
extra charge parameter $\delta$. The metrics are \cite{Cvetic:96}
 \bequ \barr{c}
\displaystyle{ds^2=-\frac{R^2(R^2-2m)}{\Sigma^2}dt^2
+\frac{R^2\Sigma dR^2}{R^4-2m(R^2-a^2)}+\frac{\Sigma}{4}\left((\sigma^1_{L,R})^2+(\sigma^2_{L,R})^2+(\sigma^3_{L,R})^2\right) } \spa{0.5}\\
\displaystyle{-\frac{2ma}{\Sigma^2}\left[R^2
(\mp\cosh^3\delta -\sinh^3\delta)+2m\sinh^3\delta\right]\sigma_ {L,R}^3 dt} \spa{0.6}\\
\displaystyle{+\frac{ma^2}{\Sigma^2}\left[\frac{R^2}{2}-m\sinh^3\delta\left\{2\sinh^3\delta\pm
2\cosh^3\delta+3\sinh\delta\right\}\right](\sigma_ {L,R})^2 \ }
\earr \label{KerrNab} \ ,\eequ where we defined \bequ \Sigma\equiv
R^2+2m\sinh^2\delta \ . \label{sigma} \eequ The gauge potential is
\bequ
A=\mp\frac{\sqrt{3}m\cosh{\delta}\sinh{\delta}}{\Sigma}\left[dt+\frac{a}{2}(\sinh\delta\pm
\cosh\delta)\sigma_{L,R}^3\right] \ \label{KNA} . \eequ The left
and right solutions are now related by \bequ a\rightarrow -a \ , \
\ \ \ \phi\rightarrow -\phi \ , \ \ \ \ \delta\rightarrow -\delta
\ . \eequ
The physical mass, charge and angular momentum for the Left Kerr-Newman are
\bequ
M=3m\cosh{(2\delta)} \ , \ \ \ \ \ Q=m\sinh{(2\delta)} \ , \ \ \ \ \ J_L=8ma(\cosh^3{\delta}-\sinh{\delta}) \ , \ \ \ \ \ J_R=0 \ . 
\eequ
This geometry has a curvature singularity at $\Sigma=0$; for sufficiently small $a$ there is an 
outer event horizon at
\bequ
R^2=m+\sqrt{m^2-2ma^2} \ . \eequ
A special case of this solution is the extremal
limit: \bequ a,b,m \rightarrow 0 \ \ \ \ \delta\rightarrow -\infty
, \label{lim1} \eequ keeping fixed \bequ me^{-2\delta}/2 \equiv \mu
\ , \ \ \ ae^{-\delta}/2\equiv \omega \ . \label{lim2} \eequ In this
limit \bequ \Sigma\rightarrow R^2+\mu \ ,\eequ and the Left
Kerr-Newman becomes \bequ \barr{c} \displaystyle{
ds^2=-\left(1+\frac{\mu}{R^2}\right) ^{-2}\left[dt+\frac{\mu
\omega}{R^2}\sigma_L^3\right]^2
+\left(1+\frac{\mu}{R^2}\right)\left[dR^2+\frac{R^2}{4}((\sigma^1_L)^2+(\sigma^2_L)^2+
(\sigma^3_L)^2)\right]} \ , \spa{0.5}\\
\displaystyle{A = \frac{\sqrt{3}}{2}\left(1+\frac{\mu}{R^2}\right)
^{-1} \left[dt+\frac{\mu \omega}{R^2}\sigma_L^3\right]} \ . \earr
\label{BMPV} \eequ This is the BMPV solution (first derived in
\cite{Breckenridge:96}), in isotropic coordinates. The right
Kerr-Newman yields, in this extremal limit, (\ref{BMPV}) with $\omega=0$. Thus, solely the
Left angular momentum survives in this extremal limit. If instead
we would have taken the $\delta\rightarrow \infty$ in the extremal
limit (\ref{lim1}) with the associated modifications for the fixed quantities,
 we would have found a BMPV solution from the Right
Kerr-Newman; only Right angular momentum survives in such extremal
limit.

The maximally
supersymmetric G\"odel type universe \cite{Gauntlett:2002nw} has
$J_R=0$. This latter solution reads \bequ
ds^2=-[dt+jR^2\sigma^3_L]^2+ds^2({\mathbb{R}}^4) \ , \ \ \ \ \
A=\frac{\sqrt{3}}{2}jR^2\sigma^3_L \ . \label{godel} \eequ
Gimon and Hashimoto showed that we can embed a Kerr black hole in this homogeneous space \cite{Gimon:2003ms}. Naturally,
the simplest case is to embed the `Left Kerr'. Then the Kerr-G\"odel solution reads \bequ \barr{l} \displaystyle{
ds^2=-\left[dt+jR^2\sigma^3_L\right]^2+\frac{R^4dR^2}{R^4-2m(R^2-a^2)+8jmR^2(a+2jm)}+\frac{R^2}{4}\left((\sigma^1_L)^2+(\sigma^2_L)^2+
(\sigma^3_L)^2\right)}\spa{0.5}\\
\displaystyle{~~~~~~~+\frac{2m}{R^2}\left[dt-\frac{a}{2}\sigma_L^3\right]^2-2mj^2R^2(\sigma_L^3)^2}\ , \spa{0.6}\\
\displaystyle{A=\frac{\sqrt{3}}{2}jR^2\sigma^3_L} \ .
\label{kerrgodel} \earr \eequ
This solution has a curvature singularity at $R=0$, which, at least for sufficiently small $a$ and $j$ is hidden behind an
outer event horizon at
\bequ
R^2=m-4jm(a+2jm)+\sqrt{[m-4jm(a+2jm)]^2-2ma^2} \ . \eequ

\sect{The Kerr-Newman-G\"odel solution} Consider the following six
dimensional truncation of the heterotic string low energy
effective field theory (in the string frame) \bequ
{\mathcal{S}}_{Het}=\int
d^6x\sqrt{-G}e^{-\phi}\left(R+\partial_{\mu}\phi\partial^{\mu}\phi-\frac{1}{12}\tilde{H}_{\mu
\alpha \beta}\tilde{H}^{\mu \alpha
\beta}-\frac{1}{8}F_{\mu\nu}F^{\mu \nu}\right) \ .
\label{hete6}\eequ The fundamental fields are the string frame
metric $G$, the NS field $B$, one abelian gauge field $A$ and the
dilaton $\phi$. The fields strengths present in the action are
related with the potentials by \bequ F=dA \ , \ \ \ \ \ H=dB \ , \
\ \ \ \ \tilde{H}=H-\frac{1}{4}A\wedge F \ . \eequ Here,
$\tilde{H}$ and $F$ are the gauge invariant objects; the gauge
transformations are the standard one for the $B$ field,
$B^{(2)}\rightarrow B^{(2)}+d\Lambda^{(1)}$, but for the $A$ field
we need \bequ A^{(1)}\rightarrow A^{(1)}+d\Lambda^{(0)} \ , \ \ \
\ \ and \ \ \ \ \ B^{(2)}\rightarrow B^{(2)}+\Lambda^{(0)}F^{(2)}
\ . \eequ We have written the rank of the form as a superscript.
The Bianchi identities are \bequ dF=0 \ , \ \ \ \ \ \
d\tilde{H}=-\frac{1}{4}F\wedge F \ . \eequ The following theorem
provides a way of generating new solutions (charged) of this theory from
known ones (uncharged):
\spa{0.2}\\
\textit{Theorem (Hassan-Sen)}\cite{Hassan:91}: Let $(G,B,A,\phi)$
be a solution of (\ref{hete6}) which is independent of time, $t$,
and a set of spatial coordinates, $x^i$. Denote $a=0,i$. Denote
the $(i+1)$ dimensional square matrices $G_ {ab}$ and $B_ {ab}$ by
$\hat{G}$ and $\hat{B}$. Denote the $(i+1)$ dimensional column
vector $A_a$ by $\hat{A}$. Assume the metric and NS field obey
$G_{\mu a}=B_ {\mu a}=0$, for $\mu\neq 0,i$. Define the Hassan-Sen
matrix as \bequ {\mathcal{M}}=\left( \barr{c}
(K^T-\eta)\hat{G}^{-1}(K-\eta) \ \ \ \ \ \ \
(K^T-\eta)\hat{G}^{-1}(K+\eta)\ \ \ \ \ \ \
-(K^T-\eta)\hat{G}^{-1}\hat{A} \spa{0.3}\\
(K^T+\eta)\hat{G}^{-1}(K-\eta) \ \ \ \ \ \ \
(K^T+\eta)\hat{G}^{-1}(K+\eta)\ \ \ \ \ \ \
-(K^T+\eta)\hat{G}^{-1}\hat{A} \spa{0.3}\\
-\hat{A}^T\hat{G}^{-1}(K-\eta) \ \ \ \ \ \ \ \ \ \ \ \
-\hat{A}^T\hat{G}^{-1}(K+\eta)\ \ \ \ \ \ \ \ \ \ \ \ \ \ \ \
\hat{A}^T\hat{G}^{-1}\hat{A} \earr \right) \ , \label{hs} \eequ
where \bequ K=-\hat{B}-\hat{G}-\frac{1}{4}\hat{A}\hat{A}^T  \ , \
\ \ \ \ \ \ \eta={\rm{diag}}(-1,1,\dots,1) \ . \eequ Then, a new,
inequivalent solution of (\ref{hete6}), $(G',B',A',\phi')$  is
obtained as \bequ \barr{c}
\displaystyle{{\mathcal{M}}'=\Omega{\mathcal{M}}\Omega^T}
\spa{0.3}\\
\displaystyle{\phi'=\phi+\ln\sqrt{\det{\hat{G}'}/\det{\hat{G}}}}
\earr \ , \label{transform}\eequ where $\Omega$ is defined as
\bequ \Omega=\left(\barr{c} S \ \ \ \spa{0.2}\\ \ \ \ R \earr
\right) \ , \label{SR} \eequ with $S,R$ being $O(1,i)$ and
$O(1,i+1)$ matrices respectively. The $S$ matrix describes Lorentz
transformations in the $t,x^i$-plane. The $R$ matrix describes
Lorentz transformations in the $t,x^i,z$ plane, where $z$ is the
internal direction associated to the gauge field $A$.

We rewrite the Hassan-Sen matrix in the notation \bequ
\mathcal{M}=\left(\barr{c} M^{--} \ \ \ \ \ \ M^{-+} \ \ \ \ \ \
X^-
\spa{0.3}\\  M^{+-} \ \ \ \ \ \ M^{++} \ \ \ \ \ \ X^+ \spa{0.3}\\
(X^-)^T \ \ \  \ \ (X^+)^T \ \ \ \ \ \ \chi \ \ \ \ \earr \right)
\label{hs2} \ . \eequ Equating (\ref{hs2}) with (\ref{hs}) defines
the matrices $M^{\pm \pm}$, the vectors $X^{\pm}$ and the scalar
$\chi$.

We can compute the fields $\hat{G},\hat{A}, \hat{B}$ in terms of
the entries of the matrix $\mathcal{M}$. The metric is computed as
\bequ
\hat{G}^{-1}=\frac{1}{4}\eta\left(M^{++}+M^{--}-M^{+-}-M^{-+}\right)\eta
\ , \label{g} \eequ the gauge field follows as \bequ
\hat{A}=\frac{1}{2}\hat{G}\eta(X^--X^+) \ , \label{A} \eequ and
the Neveu-Schwarz field can be computed from \bequ
\hat{B}=\frac{1}{4}\hat{G}\eta\left(M^{--}-M^{++}+M^{-+}-M^{+-}\right)-\hat{G}-\frac{1}{4}\hat{A}\hat{A}^T
\ . \label{B}  \eequ

\subsection{Generating the solution}
The Kerr-Newman-G\"odel solution is generated as follows:
\begin{description}
\item[i)] Uplift the Kerr-G\"odel solution (\ref{kerrgodel}) to six dimensions as a
solution of heterotic string theory (\ref{hete6}). This is
achieved by the following ansatz of the six dimensional fields
$(g,B,\phi,A)$ in terms of the five dimensional ones
$(\tilde{g},\tilde{A})$ \cite{Herdeiro:02}, \bequ \barr{c}
\displaystyle{ds^2=\tilde{g}_{\mu
\nu}dx^{\mu}dx^{\nu}+\left(dy+\frac{2}{\sqrt{3}}\tilde{A}\right)^2 \ , \ \ \ \ \phi=0 \ , \ \ \ \ \ A=0 \ ,}\spa{0.5}\\
\displaystyle{H=dB=\frac{2}{\sqrt{3}}d\tilde{A}\wedge
\left(dy+\frac{2}{\sqrt{3}}\tilde{A}\right)-\frac{2}{\sqrt{3}}\tilde{\star}d\tilde{A}
\ .}\earr \eequ Here, $y$ is the sixth dimension and
$\tilde{\star}$ is Hodge duality with respect to $\tilde{g}$. We
take the following $B$ within its gauge equivalence class \bequ
B=j(r^2-2m)dt\wedge \sigma^3_L+jr^2\sigma^3_L\wedge
dy+jml\cos{\theta}d\phi\wedge d\psi \ . \eequ Note that since the
dilaton is zero the string metric $G$ coincides with $g$.
\item[ii)] The previous solution is independent of $(t,\phi,\psi,y)$. Take $\hat{G}$ and
$\hat{B}$ to be the restriction of the six dimensional fields to
this subspace. Then, take the matrices $S$ and $R$ to be boosts in
the $y$ direction with parameters $-\alpha$ and $\alpha$
respectively \bequ S=\left(\barr{c} \cosh{\alpha} \ \ \ \  \
~~~-\sinh{\alpha}\spa{0.05}\\ I_2 \spa{0.05}\\
-\sinh{\alpha}~~~\ \ \ \ \ \cosh{\alpha} \earr  \right) \ , \ \ \
\ R=\left(\barr{c} \cosh{\alpha} \ \ \ \  \
~~~\sinh{\alpha}\spa{0.05}\\ I_2 \spa{0.05}\\
\sinh{\alpha}~~~\ \ \ \ \ \cosh{\alpha} \earr \right) \ . \eequ
$I_n$ denotes the $n$-dimensional identity. We have omitted the
last row and column of $R$ corresponding to the internal $z$
direction; in this transformation nothing happens in such
direction. A solution with non-trivial $G,B$ and $\phi$ is
obtained. The gauge field $A$ is still zero.

\item[iii)] Perform a strong-weak coupling string duality between
toroidally compactified heterotic string theory and type IIA
string theory on $K3$. The latter has string frame action \bequ
{\mathcal{S}}_{IIA}=\int
d^6x\sqrt{-G'}\left\{e^{-\phi'}\left(R'+\partial_{\mu}\phi'\partial^{\mu}\phi'-\frac{1}{12}H'^2\right)-
\frac{1}{8}F'^2+\frac{1}{2^5}\epsilon^{\mu_1\dots\mu_6}B'_{\mu_1\mu_2}F'_{\mu_3\mu_4}F'_{\mu_5\mu_6}\right\}
\ . \label{IIA6}\eequ More specifically, the IIA fields (primed)
are related to the heterotic fields (unprimed) by
\cite{Breckenridge:96} \bequ \barr{c} \phi'=-\phi \ , \ \ \ \ \
\displaystyle{G'=e^{-\phi}G}\ , \ \ \ \ \ \  H'=e^{-\phi}\star
\tilde{H} \ , \ \ \ \ \  A'=A \ .\earr \eequ The Einstein frame
metric $g=\exp{(-\phi/2)}G$ is invariant in the process. The
duality swaps equation of motion and Bianchi identity in going
from $\tilde{H}$ to $H'$. In particular the Bianchi identity
becomes the standard $dH'=0$ which means $H'=dB'$. The idea  of
the transformation is to swap the sign of the dilaton which we
will be able to cancel with another boost. Note that the
difference between these truncations of the type IIA and heterotic
actions is at the level of the gauge field $A(A')$. Since at this
point this field is vanishing, the final solution is also a
solution of six dimensional heterotic string theory.
\item[iv)]Perform a second Hassan-Sen boost, with parameter $\beta$,
this time along the internal direction $z$. Thus take, $S=I_4$ and
$R$ to be \bequ R=\left(\barr{c} \cosh{\beta} \ \ \ \  \
~~~\sinh{\beta}\spa{0.05}\\ I_3 \spa{0.05}\\
\sinh{\beta}~~~\ \ \ \ \ \cosh{\beta} \earr \right) \ . \eequ By
choosing $\beta=2\alpha$, the resulting solution has vanishing
dilaton and $g_{yy}=1$. But there is a non-trivial $B$ and $A$
fields.
\item[v)] Perform again string-string duality to obtain a type IIA solution, as in iii); Since
now the dilaton is constant, the only non-trivial change is for
the NS field. The type IIA field is \bequ H'=\star \tilde{H} \ .
\eequ This defines the IIA NS field as $H'=dB'$.
\end{description}
The resulting solution has the following six dimensional fields
\bequ
ds'^2_6=ds^2_5+\left\{dy+j\left[2m\sinh{\delta}+R^2(\cosh{\delta}-\sinh{\delta})\right]
\sigma_L^3\right\}^2 \ , \label{generalg6} \eequ \bequ \barr{l}\spa{0.02}\\
\displaystyle{A'=\frac{4m\sinh{\delta}\cosh{\delta}}{\Sigma}\left\{dt+\frac{1}{2}\left(a+2jR^2\right)
(\sinh{\delta}-\cosh{\delta})\sigma_L^3\right\}} \ ,
\label{generalA} \earr \eequ \bequ \barr{c}\spa{0.02}\\
\displaystyle{B'=\frac{jR^2(R^2-2m)(\cosh{\delta}-\sinh{\delta})}{\Sigma}dt\wedge
\sigma_L^3+jm\cos{\theta}(a-4jm\sinh^2{\delta})d\phi\wedge d\psi}
\spa{0.3}\\
\displaystyle{+\frac{m\sinh{2\delta}dt+\left[4jm^2\sinh^3{\delta}+(\cosh{\delta}
-\sinh{\delta})[jR^4+4mjR^2\sinh^2{\delta}-ma\sinh{2\delta}/2]\right]\sigma_L^3}{\Sigma}\wedge
dy} \ . \label{generalB} \earr \eequ The function $\Sigma$ is the
same as in (\ref{sigma}). We have separated the five dimensional
piece of the metric which will in fact be the final result for the
metric. It can be written as \bequ \barr{l}
\displaystyle{ds_5^2=-\frac{R^2(R^2-2m)}{\Sigma^2}dt^2+\frac{R^2\Sigma dR^2}{R^4-2m(R^2-a^2)+8jmR^2(a+2jm)}+\frac{\Sigma}{4}\left[(\sigma^1_L)^2+(\sigma^2_L)^2+(\sigma^3_L)^2\right]} \spa{0.5}\\
\displaystyle{-\frac{2}{\Sigma^2}\left\{jR^2\left[(\cosh{\delta}-\sinh{\delta})(\Sigma^2-m^2\sinh^2{(2\delta)})+2m\Sigma\sinh{\delta}\right]\right.}\spa{0.5}\\ \displaystyle{\left. ~~~~~~~+ma\left[(\cosh^3{\delta}-\sinh^3{\delta})R^2+2m\sinh^3\delta\right]\right\}dt\sigma^3_L~~~~~~~~~~~~~~~~~~~~~~~~~~~~~~~~~~~~~~~~~~~~~~~~~~~~~~}\spa{0.5}\\ \displaystyle{-\frac{m(a+2jR^2)}{2\Sigma^2}\left\{2jR^4(1-\sinh{2\delta}+4\cosh{\delta}\sinh^2{\delta}(\cosh{\delta}-\sinh{\delta}))+16jm^2\sinh^5{\delta}(\cosh{\delta}-\sinh{\delta})\right.}\spa{0.5}\\
\displaystyle{\left.~~~~~~~~~~~~~~~~~~~+4jmR^2\sinh^2{\delta}\left(2+5\sinh^2{\delta}-\sinh{2\delta}-6\sinh^3{\delta}(\cosh{\delta}-\sinh{\delta})\right)\right.~~}\spa{0.5}\\
\displaystyle{\left.~~~~~~~~~~~~~~~~~~~-a\left[R^2-2m\sinh^3{\delta}(2\sinh^3{\delta}-2\cosh^3{\delta}+3\sinh{\delta})\right]\right\}(\sigma_L^3)^2}\spa{0.5}\\
\displaystyle{-j^2\left[2m\sinh{\delta}+R^2(\cosh{\delta}-\sinh{\delta})\right]^2
(\sigma_L^3)^2} \ ,   \label{generalg5} \earr \eequ The fields
(\ref{generalg6})-(\ref{generalB}), together with $\phi=0$ can be
verified to solve the equations of motion that follow from
(\ref{IIA6}). For completeness we write down these equations (in
the Einstein frame) \bequ R'_{\mu
\nu}=\frac{1}{4}\left\{\partial_{\mu}\phi'\partial_{\nu}\phi'+e^{-\phi'}\left[H'_{\mu\alpha\beta}H_{\nu}'^{
\ \alpha \beta}-\frac{1}{6}g'_{\mu
\nu}H'^2\right]+e^{\phi'/2}\left[F'_{\mu\alpha}F_{\nu}'^{ \
\alpha}-\frac{1}{8}g'_{\mu \nu}F'^2\right]\right\} \ ,
\label{eqmot1} \eequ \bequ D_{\mu}\left(e^{-\phi'}H'^{\mu \alpha
\beta }\right)=-\frac{1}{16}\tilde{\epsilon}^{\alpha \beta
\mu_1\dots\mu_4}F'_{\mu_1\mu_2}F'_{\mu_3\mu_4} \ , \ \ \ \ \
D_{\mu}\left(e^{\phi'/2}F'^{\mu
\nu}\right)=-\frac{1}{12}\tilde{\epsilon}^{\nu
\mu_1\dots\mu_5}F'_{\mu_1 \mu_2}H'^{\mu_3\mu_4\mu_5} \ ,
\label{eqmot2} \eequ \bequ \Box
\phi'=-\frac{1}{6}e^{-\phi'}H'^2+\frac{1}{8}e^{\phi'/2}F'^2 \ .
\label{eqmot3} \eequ

\begin{description}

\item[vi)] Perform Kaluza-Klein reduction along the $y$ direction
to five dimensions. Besides the metric, $ds^2_5$ we obtain a five
dimensional NS field, $B_5$, and three gauge fields, $A'$ already
present in 6 dimensions, $A_g$ from the metric and $A_B$ from the
NS field. The KK ansatz for the metric and NS field are the
standard ones \bequ ds'^2_6=ds^2_5+[dy+A_g]^2 \ , \ \ \ \ \ \
B'=B_5+A_B\wedge dy \ . \eequ
\end{description}
The final five dimensional fields are therefore $ds^2_5$, $A'$,
$B_5$, $A_g$ and $A_B$, which are a solution to IIA string theory
on $K3\times S^1$. Note that $A'$ is the Ramond-Ramond one-form.

For $j=0$, the metric (\ref{generalg5}) reduces to
(\ref{KerrNab}), the gauge fields $A'$ and $A_B$ reduce to
(\ref{KNA}) (up to a redefinition by a constant) and $A_g$ and
$B_5$ vanish. Thus we recover the Kerr-Newman black hole. For
$\delta=0$, (\ref{generalg5}) reduces to (\ref{kerrgodel}), $A'$
vanishes, $A_B$ and $A_g$ reduce to (\ref{kerrgodel}) (up to
constant) and $B_5$ contains the same information as $A_g$ via
\bequ
dB_5=A_g\wedge A_g+\star dA_g \ .
\label{dual}
\eequ
Thus we recover the Kerr-G\"odel
black hole. In this sense we dub our five dimensional solution as
the Kerr-Newman-G\"odel black hole. We will postpone for somewhere else a more detailed 
analysis of the full solution.

\sect{Extremal limit} The extremal limit of the Kerr-Newman was
taken in (\ref{lim1}) and (\ref{lim2}). Now we have an extra
parameter, $j$. Consider first (\ref{generalg5}) with $m=0$. We
find (\ref{godel}) with $j$ replaced by
$j(\cosh{\delta}-\sinh{\delta})$. Thus, to get an asymptotic
G\"odel solution we take the following extremal limit of our five
dimensional solution: \bequ \delta \rightarrow -\infty \ , \ \ \ \
\ \ \ j,m,a \rightarrow 0 \ , \label{lim3} \eequ keeping fixed
\bequ J\equiv je^{-\delta} \ , \ \ \ \ \mu \equiv me^{-2\delta}/2
\ , \ \ \ \ \omega \equiv ae^{-\delta}/2 \ . \eequ The metric we
obtain is \bequ \barr{c}
\displaystyle{ds^2_5=-\left(1+\frac{\mu}{R^2}\right)^{-2}\left[dt+\left(JR^2+2\mu
J+\frac{\mu \omega}{R^2}\right)\sigma_L^3\right]^2}\spa{0.4}\\
\displaystyle{ +\left(1+\frac{\mu}{R^2}\right)
\left[dR^2+\frac{R^2}{4}\left((\sigma^1_{L})^2+(\sigma^2_{L})^2+(\sigma^3_{L})^2\right)\right]}
 \ , \earr \label{newBMPV} \eequ and the form fields are \bequ
A'=2\mu\frac{ -dt+(\omega+JR^2)\sigma_L^3}{R^2+\mu} \ , \ \ \
A_g=JR^2\sigma_L^3 \ , \ \ \ A_B=\frac{-\mu dt+(JR^4+2\mu
R^2J+\mu\omega)\sigma_L^3}{R^2+\mu} \ , \eequ and \bequ
B_5=\frac{JR^4}{R^2+\mu}dt\wedge \sigma_L^3 \ . \eequ
For $J=0$, (\ref{newBMPV}) reduces to (\ref{BMPV}), $B_5$ and $A_g$ vanish and $A'$ and $A_B$ give (\ref{BMPV}) up to a 
gauge transformation and an overall factor. For $\mu=0$ (\ref{newBMPV}) reduces to (\ref{godel}), $A'$ vanishes, $A_g=A_B$ 
reduce to (\ref{godel}) up to constant and $B_5$ is related to $A_g$ by (\ref{dual}). The general solution has a coordinate 
singularity at $R^2=0$. For $4\omega^2/\mu<1$, this surface is a null surface and 
can be interpreted as the black hole event horizon, exactly as for the usual (asymptotically flat) BMPV black hole.
This is easily  seen by changing to a Schwarzschild like radial coordinate defined by $r^2=R^2+\mu$; the coefficient of $\sigma_L^3$ in the metric becomes
\bequ
\frac{r^2}{4}-\frac{(Jr^4-J\mu^2+\mu\omega)^2}{r^4} \ , \eequ
and thus the J dependence vanishes at $r^2=\mu$. More precisely one can introduce a set of regular coordinates on the horizon following \cite{Herdeiro:02} and 
show that for $4\omega^2/\mu<1$, $r^2=\mu$ is a null surface.
The black hole entropy is then
\bequ
S=\frac{\pi^2\mu}{2G_5}\sqrt{\mu-4\omega^2} \ . \eequ

There is a true physical singularity (timelike and point-like) at $r^2=0$, since the Ricci scalar diverges there, as $\mu^3/r^8$.

\sect{Conclusions}
In this letter we have derived a new solution to type II string theory in five dimensions, 
interpreted as a Kerr-Newman black hole in a G\"odel type universe. This was achieved by a set of Hassan-Sen transformations
and string dualities. Recently \cite{Elvang:2003yy}, a Hassan-Sen transformation was used to give charge to the rotating black ring of Emparan and Reall \cite{Emparan:2001wn}; this ring is another surprise of five dimensional gravity, with no four dimensional equivalent, as the BMPV black hole and the G\"odel type
universe  discussed herein. Perhaps the five dimensional Petrov classification \cite{ DeSmet:2002fv,DeSmet:2003rf} will be a useful tool in understanding and classifying 
these various solutions.

 The most interesting special case of our new solution is an extremal limit which is BPS. The BMPV black hole has a ten dimensional
interpretation as a D1-D5-Brinkmann wave system \cite{Herdeiro:00}; the G\"odel spacetime has a ten dimensional interpretation as a pp-wave \cite{Boyda:2002ba}. It will therefore 
be interesting to understand in more detail the ten dimensional interpretation of the BMPV-G\"odel solution, in particular in relation to the issue of existence of horizons in pp-waves \cite{Gimon:2003xk, Hubeny:2002pj,Liu:2003ct}.

\section *{Acknowledgments}
The author is supported by the grant SFRH/BPD/5544/2001 (Portugal). This project is also supported by POCTI/FNU/38004/2001-FEDER.

\end{document}